\documentclass[sigconf, natbib=true, anonymous=false]{acmart}

\AtBeginDocument{%
  }

\setcopyright{acmlicensed}
\copyrightyear{2024}
\acmYear{2024}

\usepackage{gensymb}

\acmConference[SIGIR]{}{July 18, 2024}{Washington D.C., USA}
\begin{document}

\title[IR in Climate Downscaling]{Identifying high resolution benchmark data needs and Novel data-driven methodologies for Climate Downscaling}

\author{Declan Curran}
\orcid{1234-5678-9012}
\affiliation{%
  \institution{University of New South Wales}
  \city{Sydney}
  \state{New South Wales}
  \country{Australia}
}
\email{d.curran@unsw.edu.au}

\author{Hira Saleem}
\affiliation{%
  \institution{University of New South Wales}
  \city{Sydney}
  \state{New South Wales}
  \country{Australia}
}
\email{h.saleem@unsw.edu.au}

\author{Flora D. Salim}
\affiliation{%
  \institution{University of New South Wales}
  \city{Sydney}
  \state{New South Wales}
  \country{Australia}
}
\email{flora.salim@unsw.edu.au}

\renewcommand{\shortauthors}{Curran et al.}
\begin{abstract}
We address the essential role of information retrieval in enhancing climate downscaling, focusing on the need for high-resolution datasets and the application of deep learning models. We explore the requirements for acquiring detailed spatial and temporal climate data, crucial for accurate local forecasts, and discuss how deep learning (DL) techniques can significantly improve downscaling precision by modelling the complex relationships between climate variables. Additionally, we examine the specific challenges related to the retrieval of relevant climatic data, emphasizing methods for efficient data extraction and utilization to support advanced model training. This research underscores an integrated approach, combining information retrieval, deep learning, and climate science to refine the process of climate downscaling, aiming to produce more accurate and actionable local climate projections.
\end{abstract}

\maketitle

\section{Introduction}
Climate model predictions are usually only available at very low resolutions due to the severe computational requirements involved in running these models\cite{CMIP6}. Nonetheless, high resolution data is important for extreme weather event preparation, and climate preparation in general \cite{EnhancingRegionalClimateDownscalingthroughAdvancesinMachineLearning}. The same can be said for select meteorological variables such as soil moisture content and sea surface temperature which both have large impacts on earth’s weather patterns, but are only available at low level resolutions in some sources \cite{doi:10.1080/01431161.2017.1356486,8668494}. Climate downscaling is the process of taking these low-resolution measurements and producing high resolution data \cite{SUN202414}. 

Climate downscaling is split into two fields: dynamical downscaling and statistical downscaling. Dynamical downscaling involves running climate models at finer resolutions than the usual 100km by 100km grid but can normally only be done over small areas due to resource constraints \cite{Gao2022}. Statistical downscaling involves finding proxies in weather variables and using this to aid in predicting higher resolution results—this subfield has exploded in popularity through recent years due to the advent of improved deep learning (DL) models and forms the subject of this review \cite{SUN202414}.

There has been a wide volume of research into statistical downscaling through a variety of ML methods and data sources \cite{Hassan2015,SACHINDRA2018240,Goutham2021}. Yet the field has been marked by two common driving forces over the past few years which present exciting opportunities for future innovation: more high resolution training data and improvements in the scaling abilities of new DL models. 

Graphcast and Panguweather are two recent DL models that have been applied to the weather forecasting domain—being the first to beat the widespread incumbent Numerical Weather Prediction (NWP) models \cite{bi2023accurate,doi:10.1126/science.adi2336}.

Similar weather foundation models have seen some success in downscaling \cite{nguyen2023climax}. However, the benefits of new scalable methods—such as transformers and graph attention networks—have not yet been fully realised. This is partly due to the aforementioned lack of available high resolution datasets historically and the intense computational requirement to use this data at scale. 

Currently, downscaling resolution is limited by dataset size for supervised methods—of which the highest resolution datasets are normally at a resolution of several kilometres \cite{Su2018,Powers2017TheWR,NARO}. These have facilitated several downscaling studies which have pushed the field to the limits of its current resolution \cite{mardani2023residual,Damiani2024}. Some studies have combined satellite datasets which are capable of achieving higher resolutions—even reaching the meter level—but there is still work to be done in this area as discussed further in section 2. 

\section{Dataset Requirements in Climate Downscaling}
The greatest barrier to current downscaling resolution is the availability of very high resolution datasets. ERA5 reanalysis data is the gold standard for global meteorological data; however, it's main product is limited to 0.25 degree data---which is not granular enough for specialised climate preparation scenarios \cite{EnhancingRegionalClimateDownscalingthroughAdvancesinMachineLearning}. 

Unfortunately benchmarks in this area have historically been few and far between. The increased pace of innovation in climate science has meant that this is changing with weatherbench and climateSET for weather specific datasets and even downscaling specific datasets that incorporate need for high resolution data \cite{linTangDeep,Languth}.  The prevalence of high resolution region-specific datasets has also increased in recent years which have proved valuable in bringing downscaling measurements to several km wide \cite{mardani2023residual,Damiani2024}.

\textbf{Reanalysis Datasets.}
The ERA5 dataset, provided by the European Centre for Medium-Range Weather Forecasts (ECMWF) \cite{Hersbach2020-dq}, represents one of the most advanced atmospheric reanalysis of the global climate available today. ERA5 provides hourly estimates of a vast range of atmospheric, land-surface, and sea-state parameters. The spatial resolution is approximately 31 km (0.25 degrees on a latitude-longitude grid) on 137 vertical levels from the surface up to a height of 80 km. The temporal resolution is hourly, offering a more detailed representation of diurnal cycles compared to ERA-Interim. It covers data from 1950 onwards, allowing for historical climate analysis and providing a valuable tool for understanding long-term climate trends. ERA5 incorporates vast amounts of historical observational data, assimilated into the climate model to produce what is intended to be a highly accurate representation of the historical state of the Earth's atmosphere.

\textbf{Climate Datasets.}
Climate models rely on detailed assumptions and each individual climate model may be tailored to a specific stream of climate literature \cite{article}. To assist with standardisation and comparison of results, GCMs are run on a standard set of simulations administered by the Coupled Model Intercomparison Project (CMIP) \cite{article}. CMIP6 is the latest iteration of climate scenarios and represents the culmination of work from 49 different groups across hundreds of climate models [22].  

\textbf{Benchmark Datasets.}
Benchmark datasets provide an important role in standardising input data amongst models to better track model performance and allow for easier data collation. While several benchmarks such as ClimSim \cite{yu2024climsim}, ClimateSet \cite{kaltenborn2023climateset} and ChaosBench \cite{nathaniel2024chaosbench} have emerged for climate specific tasks including downscaling in recent years, they are still limited in their spatial and temporal capabilities. As such, a standardised benchmark has yet to be adopted in the field of downscaling.

\textbf{Region-Specific Datasets.} 
Region-specific datasets play an important role in current downscaling research by allowing for high resolution output data. This data comes in several formats, with NWP forecasts (such as RWF model outputs) forming valuable training data for some models \cite{mardani2023residual,https://doi.org/10.48550/arxiv.2208.05424,Powers2017TheWR}. Reanalysis data is also available in some regions which combines predictions with real observed data \cite{Su2018,NARO}. Datasets like this are critical for further improvements in downscaling resolution.



\textbf{Satellite Data.}
Satellite measurements provide a powerful measure for collecting meteorological data and have been used in conjunction with weather variables to great success in some downscaling tasks  \cite{SAGAN2020103187,Kim2023,Zhang2024,Xu2021}. The value of this information has necessitated novel methods to incorporate such data but this has nonetheless proven difficult; satellite data is usually available at a much higher spatial resolution than reanalysis data but longer temporal frequency—having to complete several rotations of the earth to return to the same location unless the satellite is in geostationary orbit. 

Regardless, this area still remains relatively under explored in downscaling. 

\begin{table}[!htbp]
  \caption{Examples of Satellite Datasets in Downscaling}
  \label{tab:freq}
  \begin{tabular}{cccc} 
    \toprule
    Satellite&Resolution& Revisit Time & Variable\\
    \midrule
    Sentinel-2 [45]& 10m x 10m& 5 days & Images\\
    G11SST [42]& 1km x 1km& Daily & Sea Surface Temp\\
    Landsat [45]& 30m x 30m & 8 days& Images\\
    MOD13Q1 [44] & 250m x 250m & 16 days & Vegetation\\
  \bottomrule
\end{tabular}
\end{table}


\begin{figure}[h]
  \centering
  \fbox{\includegraphics[width=.5\textwidth]{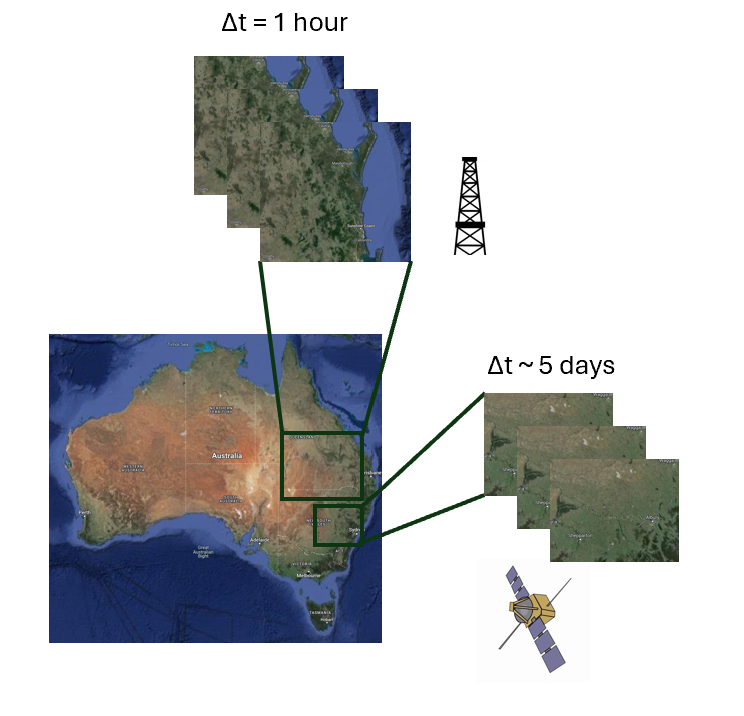}}
  \caption{Datasources with varying spatial and temporal resolution can be combined to form composites that have high spatial and temporal resolution. ERA5 data is available at ~31km resolution at 1 hour intervals whereas sentinel-2 satellite data is available at a 10m resolution every 5 days on average over a given area of earth.}
  \Description{Data Fusion}
\end{figure}

\textbf{Need for a High Resolution Benchmark Dataset.}
Regional climate models suffer from several known limitations involving an underestimation of extreme events; this can be linked to the way they model convection \cite{Fosser2024}. Convection is a meteorological process whereby heat and moisture travel vertically within the atmosphere and has a large effect on the formation of storms and other extreme events \cite{https://doi.org/10.1002/2014JD021478}. Modelled results larger than a 4km resolution are known to be unable to approximate convection well; particularly in areas with complex topography—consequently, models predicting data at a resolution between 4km and 1km are known as ‘convection permitting models’ where they can approximate convection over large areas \cite{Adinolfi2023}. Dynamical downscaling methods have been popular to apply in this area for many years, with impressive results \cite{Coppola2018,Kendon2023,Prein2015} but have been limited in achieving full convection modelling, as a prohibitive 100m resolution or higher required to do this effectively \cite{Adinolfi2023}.

This is where we return to statistical downscaling, and the favourable scaling properties of current DL methods. Although satellite meteorological data has previously been used to address downscaling problems, the literature has yet to include high resolution satellite imagery---which is available at levels far below the 100m resolution required for modelling convection\cite{Sentinel}—--as a model parameter for downscaling climate models. Although this data may not be as informative as direct meteorological measurements, transformers have nonetheless been shown to be uniquely skilled at modelling similar multi-modal data problems in other domains \cite{xu2023multimodal}. We believe this is one of the next innovations in the field which will further improve results.

In addition to this, higher resolutions permit models to better understand and predict underlying heterogeneities due to the higher spatial resolution, further improving the model. Achieving this milestone could revolutionise the way in which extreme events are predicted and modelled, providing massive benefits to society.

\section{MODEL REQUIREMENTS}
\textbf{Current Methods.} The statistical downscaling problem can be formulated in a straightforward manner. We aim to find the function which approximates high resolution meteorological data $X_{HR}$ based on $X_{LR}$ and an optional $T_{HR}$ vector containing auxillary information which is usually at a higher resolution but does not have to be.

\begin{equation} X_{HR} = F(X_{LR},T_{HR})  \end{equation}

Note that this differs from the traditional super-resolution problem in computer vision in that additional auxillary information allows the model to better discern higher resolution relationships. Physical and environmental constraints can also be incorporated into the model to improve results \cite{https://doi.org/10.48550/arxiv.2208.05424}.

Historically, a number of different DL methods have been used to approximate this function including CNN, GAN, SVM and many more \cite{Hassan2015,SACHINDRA2018240,Goutham2021}. These have normally fallen under deterministic methods which compute one iteration of the above function. It is worth noting that predictive methods have recently emerged as a popular alternative that aim to model the whole distribution probabilistically \cite{mardani2023residual}. Predictive methods are well suited to a problem like this where we are predicting the noise by interpolating images but require more computational resources to enact multiple runs of the distribution---diffusion has been a popular model that has been applied with some success in this area \cite{mardani2023residual,https://doi.org/10.48550/arxiv.2401.05932}.

Large Language Models (LLMs) present another potential future avenue for climate downscaling. Often trained on billions of parameters, LLMs are well known for their ability to incorporate multi-modal data and perform competitively in time-series forecasting applications similar to weather-prediction \cite{https://doi.org/10.48550/arxiv.2402.01801}. Although they have not been applied directly to climate downscaling, there is an increasing prevalance of climate-specific foundation models in recent times which have achieved competitive results \cite{https://doi.org/10.48550/arxiv.2405.04285}.



\section{Web-scale knowledge for Climate Downscaling}

With over 5 billion humans accessing the internet in 2021, the online world provides an interesting and novel way to crowd source data on current events \cite{HumInt}. Particularly for climate applications, this is under-explored, despite potential to address an area that climate models often miss: the human impact. 

Moore et al. \cite{Moore2020} use social media data following major flooding events to identify potential areas where flood thresholds of nearby tide gauges may be inaccurate given the human response. Various other studies have also used social media data to link human behaviour with climate outcomes and perceptions \cite{Effrosynidis2022}. 

This data often is often of a multi-modal nature with images, text and geolocation all being captured\cite{Moore2020}. Novel climate datasets like this offer the potential for real time information to feedback into a model. However, it is worth noting the potential bias introduced when incorporating data generated by humans; the applicability of such data to climate downscaling is also not well documented. Nonetheless, there is potential for further use of this data in weather-ML contexts, especially when combined with other objective metrics like reanalysis and satellite data.  

\section{Conclusion}
The increased pace of innovation in climate-ML has already begun to facilitate more cost-effective higher resolution climate data. With industry trends towards more higher resolution datasources and improved DL methods, we believe there is potential for higher resolution satellite imagery and open source data from the web to further improve downscaling resolution and performance. Once this breaches the threshold to model convection, these models will provide valuable tools for extreme weather event prediction and general climate preparedness. 

The paper presents an open set of challenges and problems, covering both data and model requirements for climate downscaling, and invites researchers in related fields to contribute to works in climate downscaling.




\bibliographystyle{ACM-Reference-Format}
\bibliography{ref} 

\end{document}